**Title:**
"Forming Massive Terrestrial Satellites through Binary-Exchange Capture"

**Running Title:**
"Capture of Massive Terrestrial Satellites"

**Authors:**
Darren M. Williams[1*] and Michael E. Zugger[2]
[1]Penn State Behrend, School of Science, 4205 College Drive, Erie, PA 16563
*Email: dmw145@psu.edu
*Phone: 814-898-6008
*ORCID: 0000-0002-9350-8377
[2]Penn State Applied Research Laboratory, P.O. Box 30, State College, PA 16801

**Affiliations:**

D. Williams is a member of Sigma Xi, The Scientific Research Society; the American Astronomical Society; and the following Penn State organizations:
Center of Exoplanets and Habitable Worlds (CEHW)
Planetary Systems Science Center (PSSC)
Consortium for Planetary and Exoplanetary Science and Technology (CPEST)

M. Zugger is a member of the above Penn State organizations as well as the Blue Marble Space Institute of Science (BMSIS) located in Seattle WA

**Abstract:**

The number of planetary satellites around solid objects in the inner Solar System is small either because they are difficult or unlikely to form, or that they do not survive for astronomical timescales. Here we conduct a pilot study on the possibility of satellite capture from the process of collision-less *binary-exchange* and show that massive satellites in the range $0.01 - 0.1$ $M_\oplus$ can be captured by earth-sized terrestrial planets in a way already demonstrated for larger planets both in the Solar System and possibly beyond. In this process, one of the binary objects is ejected, leaving the other object as a satellite in orbit around the planet. We specifically consider satellite capture by an 'earth' in an assortment of hypothetical encounters with large terrestrial binaries at 1 AU around the Sun. In addition, we examine the tidal evolution of captured objects and show that orbit circularization and long-term stability are possible for cases resembling the Earth-Moon system.

# 1. Introduction

Of the ~158 known natural satellites in the Solar System, only 13 orbit significant terrestrial-sized objects (500 km $< R <$ 10$^4$ km) and only three exist within 2 AU of the Sun. Most of the satellites belong to the gas-giant planets and either formed through accretion within the circumplanetary nebulae of their parent objects (Canup and Ward 2002; Ronnet and Johansen 2020; Estrada and Mosqueira 2006) or were captured. Capture involving binaries is the subject of this paper, but isolated masses can also be caught in encounters with planetary atmospheres, rings, or existing satellites (Goldreich et al. 1989; Porter and Grundy 2011). Capture to *form* binaries (mainly involving asteroids) is accomplished via *dynamical capture* in which slow-moving, weakly-interacting mass pairs are "hardened" by chance encounters with third "intruder" objects, as described by Noll et al. (2008).

Collision is thought the best mode to form binaries with large mass ratios, as is common with Main-Belt binaries (Noll 2005, Walsh and Jacobson 2015) and some Kuiper-Belt binaries such as Eris, Haumea, and Pluto-Charon (Canup 2005, Cheng et al. 2014). A similar collision is implicated in forming the Moon around the Earth (Canup and Esposito 1996; Barr 2016), but with an energy and an impact angle that completely melted and/or vaporized the smaller object. The Moon is thought to have later assembled out of the collision debris enveloping the magma Earth.

Such collisions were inevitable in the final stages of accretion leading up to planets, with many collisions ending in merger or fragmentation without forming satellites (Raymond and O'Brien 2009; Chambers 2013). But the Earth-Moon and Pluto-Charon examples, along with the ~75 known KBO binaries (Noll et al. 2008, Grishin 2020, Noll 2020) and asteroid binaries (Noll 2005, Margot et al. 2015), show that binaries do form in a variety of ways and often with remarkably large mass ratios $>$ 10:1.

We infer from these facts that a population of *binary terrestrial objects* (BTOs), perhaps exceeding 5-10% of all objects (Morbidelli et al. 2009), may have existed in the early Solar System either while (or after) the planets were achieving (achieved) their present mass. Tidal disruption of such a binary in the outer Solar System during a *binary-exchange* encounter with Neptune may have resulted in the retrograde capture of its massive satellite Triton (Agnor and Hamilton 2006), followed by subsequent circularization of its orbit by interactions with a primordial satellite system (Rufu and Canup 2017).

Here we consider whether capture and orbit circularization (albeit through tides) might apply to the Earth-Moon system. Recent geochemical analysis of lunar and chondritic material (Cano, E.J. *et al* 2020, Dauphas, N. et al. 2014) has rekindled the debate about where in the Solar System the lunar progenitor originated and how much, if any, of the progenitor was incorporated into the Moon after a collision. The possibility of collision-less capture for a 'moon' formed in the vicinity of Earth is now on the table given these updated geochemical constraints. We examine the dynamical aspects of plausible lunar capture and tidal circularization in Section *4* of this paper.

## 2. Binary-Exchange Capture

Whether binary-exchange applies to the Earth-Moon system or not, it has been shown (Williams 2013: hereafter W13) to be an effective mode of capture for satellites larger than Mercury (0.051 $M_\oplus$) or even Mars (0.107 $M_\oplus$) around gas-giant planets, provided such massive binaries exist in the first place and interplanetary encounters are common. For this study, we take the existence of BTOs in the vicinity of an 'earth' to be a given, however they might form, and reserve investigation into their origins for future work.

A key point from W13 is that smaller planets capture moons more effectively because encounter velocities within their weaker gravity wells are slower. For a successful exchange, a binary must be close enough to a planet for tidal disruption while still traveling slow enough to facilitate capture. The reader is referred to W13, specifically Figs. (1-2) and Eqns. (1-6), for a thorough discussion of encounter details. Here we focus our attention on W13 Eq. (6) (Correcting a typo in W13 by replacing *b* with *q*):

$$m_1 < 3M_p \left( \frac{G\pi m_2}{2qv_{\text{enc}}(v_{\text{enc}} - v_{\text{peri}})} \right)^{3/2} - m_2 \qquad (1)$$

with $m_1$ as the captured mass, $m_2$ as the escaping mass, $M_p$ as the planet mass, $v_{\text{enc}}$ as the velocity of the binary barycenter at closest approach, and $v_{\text{peri}}$ as the velocity of the captured mass at periapse distance *q* from the planet.

The expression in Eq. (1) is the maximum mass possibly captured with the binary components at their maximum *Hill* separation (i.e., in their most fragile state) before tidal disruption. For analytic simplicity, the derivation of Eq. (1) does not include the acceleration of the captured mass by the escaping mass after disruption.

However, including this acceleration would only *promote* capture of the new satellite by reducing its circumplanetary velocity - the pull of $m_2$ is directed away from the motion of $m_1$ after disruption - so we can safely ignore this analytic complication here.

We use Eq. (1) to compute the largest satellite possibly captured around an 'earth' at 1.0 AU from the Sun as a function of escaping mass $m_2$, approach velocity at infinity $v_\infty$, and periapse distance $q$. The captured mass $m_1$ is greatest when its circumbinary velocity is directly opposite the binary-encounter velocity at the moment of disruption, which is only approximately realized in the actual three-body encounter.

Rotational direction of the binary is also important. Figure 1 illustrates the condition that the inner, "to be captured", mass $m_1$ is the one moving opposite the binary encounter velocity. This arrangement ensures that the outer mass $m_2$ carries away the excess energy and momentum and leaves the system before significantly perturbing mass $m_1$ and compromising capture, which is the underlying assumption in the derivation of Eq. (1). Thus, while capture of a retrograde-moving outer mass is possible in principle, it cannot be modeled analytically and has been found in this study to be numerically improbable; of the dozens of runs we performed, no captures were found with the final orbital angular momentum opposite the initial direction of binary spin. Retrograde capture, then, is only likely if the binary rotates in the same direction as the captured mass $m_1$ orbits the planet.

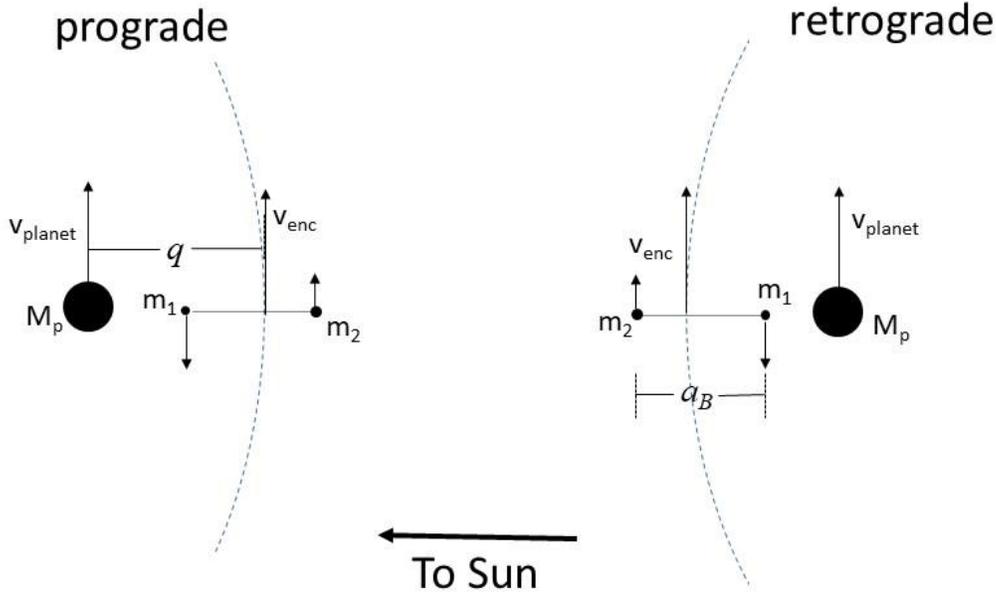

**Fig. 1**: Two-dimensional geometry for capture with a binary rotating counterclockwise (prograde) and clockwise (retrograde) relative to the planet velocity $v_{\text{planet}}$ with respect to the Sun, which is off-screen to the left. The hyperbolic encounter path of the binary barycenter around the planet is shown as a dotted line. Velocity of the binary barycenter at periapse is $v_{\text{enc}}$ and is figured in the planet reference frame. Other variables are defined in the text. Object sizes and orbits are not to scale.

After disruption, the captured mass $m_1$ enters a highly eccentric orbit with periapse distance equal to

$$q - \frac{a_B m_2}{(m_1 + m_2)} \approx q = a(1 - e), \qquad (2)$$

where $a$ and $e$ are the elements of the new planet-centric orbit. The second term on the left-hand side of Eq. (2) is the moment arm of $m_1$ within the binary and is generally much smaller than $q$ with our assumed use of $a_B = 10(R_1 + R_2)$, except in the highest mass cases considered below (e.g., when $m_1 > 0.05$ M$_\oplus$).

Equation (1) shows that the exchange reaction is independent of the choice of $a_B$. This is because the tidal acceleration ($\propto \Delta v_{\text{tidal}}$) of the inner mass ($m_1$) by the planet is $\propto a_B$, just as the height of a tide on Earth is proportional to Earth's radius $R_\oplus$, whereas the acceleration ($\propto \Delta v_{bin}$) of the inner mass by the outer mass ($m_2$) during the encounter is, according to Eq. (5) of W13,

$$\Delta v_{bin} = \left(\frac{Gm_2}{a_B^2}\right)\left(\frac{\pi q}{2v_{enc}}\right), \qquad (3)$$

where the items in () brackets are, from left to right, the acceleration of $m_1$ by $m_2$, and the approximate time over which the acceleration occurs, respectively. And since $q \propto a_B$ for tidal disruption (setting $a_B$ equal to the binary Hill-radius $r_H \propto q$), $\Delta v_{bin} \propto a_B^{-1}$, thereby offsetting the tidal acceleration and canceling the dependance on $a_B$. More succinctly, binaries with larger $a_B$ are easier to separate, but their slower rotations make capture less likely.

Here we follow W13 and impose the conservative condition that apoapse distance be less than the stability limit for massless particles on circular orbits within the circumplanetary Hill sphere. Thus,

$$a(1+e) < f_H a_p \left(\frac{M_p}{3M_*}\right)^{1/3}, \qquad (4)$$

where $M_p$ is the planet mass as before, $M_*$ is the mass of the star, and

$f_H \approx 0.5$ and $\approx 0.9$ for prograde and retrograde orbits, respectively. Retrograde orbits are more stable than prograde orbits due to the opposite sign of the coriolis acceleration in the non-inertial reference frame of the planet: *toward* the planet for retrograde-moving and *away from* the planet for prograde-moving satellites. The above condition (4) affects the size of the captured mass by limiting the periapse velocity $v_{peri}$ in the denominator of Eq. (1) which is $\propto a^{-1/2}$.

We first use Eq. (1) to compute values of captured mass over a range of escaping masses $m_2$, binary encounter velocities at infinity $v_\infty$, and periapse distances $q$. Figures 2A and 2B are the result and show the same trends found for satellite capture around gas-giant planets in W13: maximum captured mass increases with escaping mass, and tends to decrease with increasing $v_\infty$ and increasing $q$. Also, a careful inspection of Figs. 2A and 2B reveals that the captured mass limits are slightly greater for retrograde orbits than for prograde since the maximum periapse velocity of the captured mass just after tidal disruption is larger.

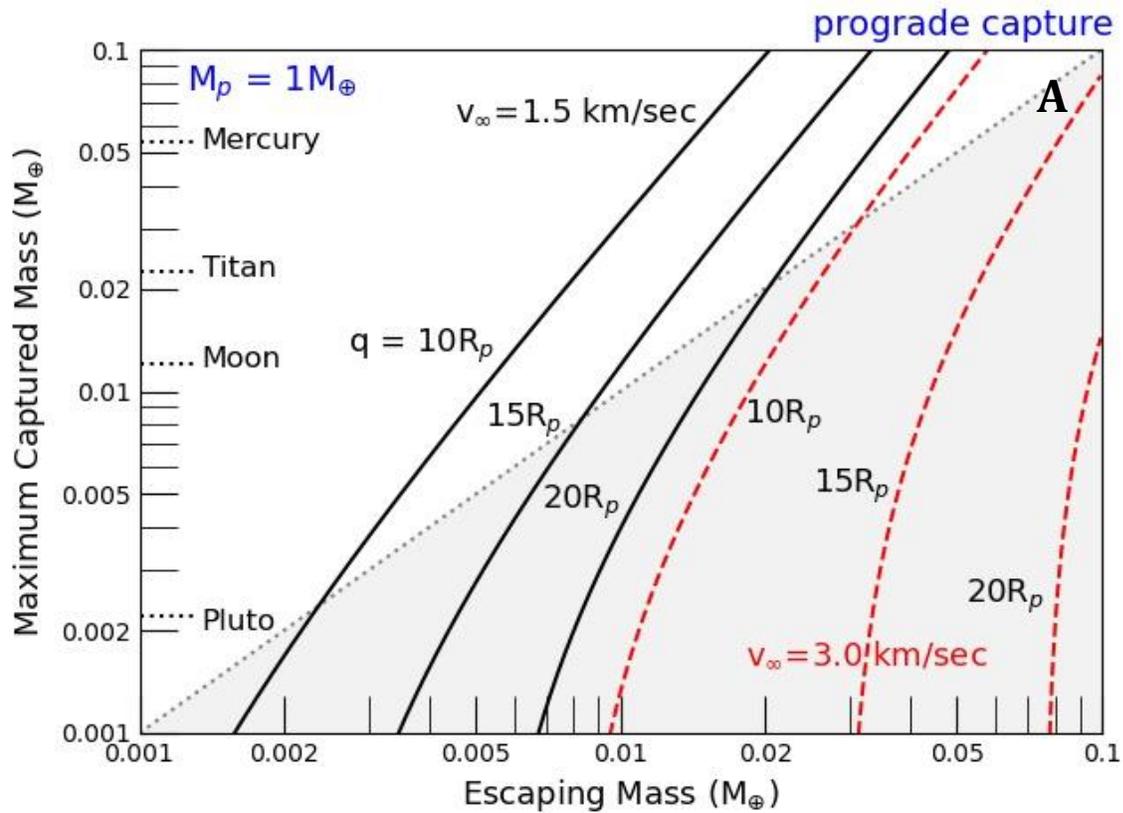

**Fig. 2:** Maximum captured mass as a function of escaping mass, minimum encounter distance ($q = 10, 15, 20\ R_p$), and binary encounter velocity at infinity ($v_\infty = 1.5, 3.0$ km sec$^{-1}$ in black solid and red dashed lines, respectively). Curves are calculated using Eq. (1) with planet mass $M_p = 1 M_\oplus$, $R_p = 1 R_\oplus$ and distance from the star $d_* = 1.0$ AU. All planet and binary objects are assumed to have the same density $\rho = 4000$ kg m$^{-3}$. Prograde capture (Fig. 2A) means that $f_H = 0.5$ in Eq. (4), whereas retrograde capture (Fig. 2B) has $f_H = 0.9$. Higher velocity ($v_\infty = 3.0$ km sec$^{-1}$) restricts capture to smaller objects in the region shaded gray, except for extremely-close encounters with $q = 10 R_p$.

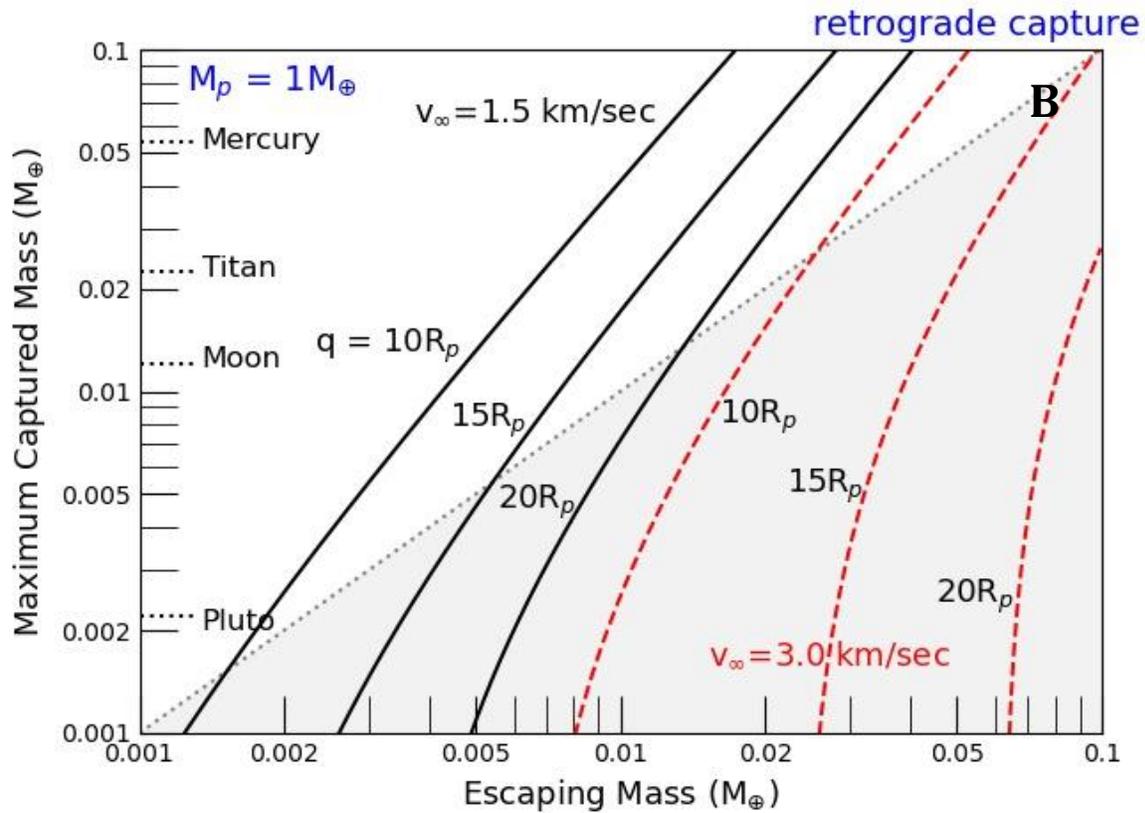

For these calculations and numerical integrations to follow, we assume that the encounter between the binary and planet occurs at 1.0 AU from a solar-mass star with a relative velocity in the range 1.5 km sec$^{-1}$ < $v_\infty$ < 3.0 km sec$^{-1}$, which is the approximate relative velocity of two objects moving in the same direction on intersecting orbits of small eccentricity separated by ~0.1 AU < $\Delta a$ < 0.2 AU. Thus, the encounters are assumed to involve objects in close proximity within the population in which they occur. In addition, the orbits of the binary and planet are assumed to be co-planar to ensure a maximum $\Delta v$ from the circulation of the binary at disruption, which is slightly greater than would be achieved in a true, randomized, three-dimensional integration. Also, the binary approach velocity must be nearly parallel (or anti-parallel) to the planet velocity to ensure a small impact parameter $b$ and an even smaller encounter distance $q$ < 20 $R_p$.

Figure 2A shows that a lunar-size mass ($m_1$ = 0.012 $M_\oplus$) could possibly be captured around Earth if the escaping companion has a mass $m_2$ > 0.001 $M_\oplus$ depending on $q$ with $v_\infty$ = 1.5 km sec$^{-1}$. Increasing the encounter velocity to 3.0 km sec$^{-1}$ makes it harder for an 'earth' to capture a lunar mass, with the required escaping mass $m_2$ > 0.02 $M_\oplus$. Thus, it is easier to capture a given mass if the relative encounter velocity of the binary is slower, as stated earlier. Switching to a retrograde approach direction in Fig. 1 shifts the curves in Fig. 2B to the left by ~70-80%, which is the reduction in mass lost from the system that is necessary to form a satellite through binary-exchange. It is therefore slightly easier to capture a satellite in a retrograde orbit than in a prograde one, so long as the retrograde capture has the binary also rotating in the same direction, as will be discussed further below.

## 3. Numerical Integration

We now take a numerical approach to test the analytical predictions represented in Fig. 2, although we do not attempt a comprehensive study of capture over the entire range of masses. Rather we focus our attention on an assortment of plausible encounters by using a simple (second-order) leap-frog, orbit integrator to track the three masses (four including the Sun) involved in a binary-exchange encounter. We tested the accuracy of our integrator against the *IAS15* routine in the *REBOUND* package (Rein and Spiegel 2015) and found only minor position discrepancies ($< 0.1$ $R_p$) after time $t \sim 0.01$ years of integration. Also, our leap-frog integrator uses an adjustable step size that diminishes linearly with object distance from the planet to conserve energy to $\Delta E/E \sim 10^{-7}$.

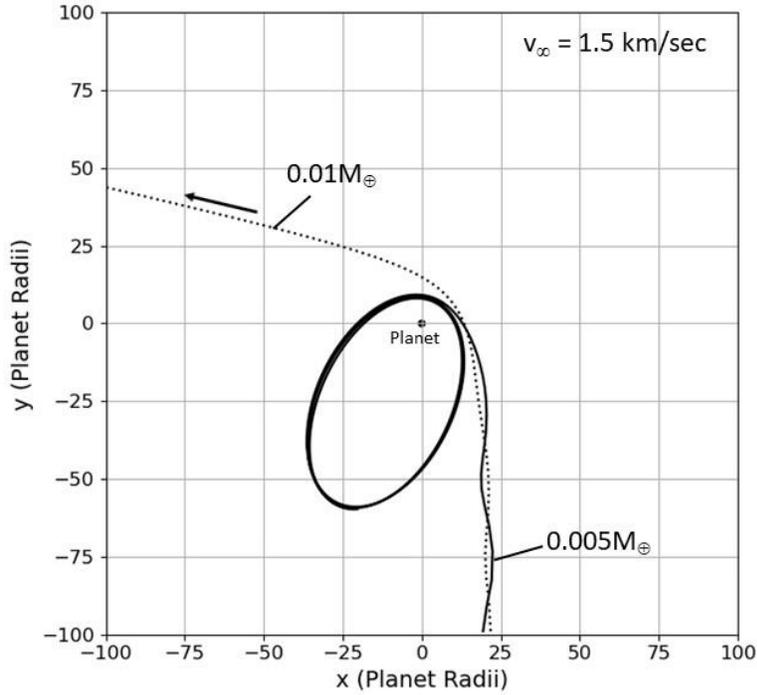

**Fig. 3:** Successful prograde capture of a satellite around an Earth-mass planet (black dot at center). The captured object (path represented with solid line) has a mass $m_1 = 0.005$ $M_\oplus$, whereas the escaping mass (dotted line) has a mass $m_2 = 0.01$ $M_\oplus$, comparable to the Moon. The initial separation of the binary components is $a_B = 10(R_1 + R_2) = 2.4$ $R_\oplus$ yielding a binary rotational period $P = 1.77$ days. The capture orbit has $q = 8.6$ $R_p$, $a = 36.1$ $R_p$, and $e = 0.76$.

---

We show a successful capture in Fig. 3 with a 0.005 $M_\oplus$ : 0.01 $M_\oplus$ binary launched from 100 $R_\oplus$ behind, and from a horizontal distance $b = 20$ $R_\oplus$ to the right of, the planet. The starting velocity at this location $v_0 = 1.87$ km/sec and is computed using

$$v_0^2 = v_\infty^2 + 2GM_p/r_0 \qquad (5)$$

where $v_\infty = 1.5$ km sec$^{-1}$ and $r_0 = 102$ $R_\oplus$ is the Pythagorean distance from the planet.

Now consider the possibility of capturing the larger mass of the same binary with all other parameters held the same. Figure 2A shows the heaviest mass that can be captured when losing a 0.005 $M_\oplus$ object (with $v_\infty = 1.5$ km sec$^{-1}$) is just under a lunar mass with $q = 10R_p$. But with the actual $q$ in Fig. 3 shown to be slightly less than this, we surmise correctly that capture of the larger mass should work if the binary is flipped at tidal disruption so that the larger mass is the one moving opposite the encounter direction. We adjust the orientation of the binary at disruption by simply adding distance (and time) to the approach path. With an average approach velocity of v ~ 1.6 km sec$^{-1}$, the added approach distance needed to flip the binary at periapse is $d = vP/2$ or ~19 $R_\oplus$.

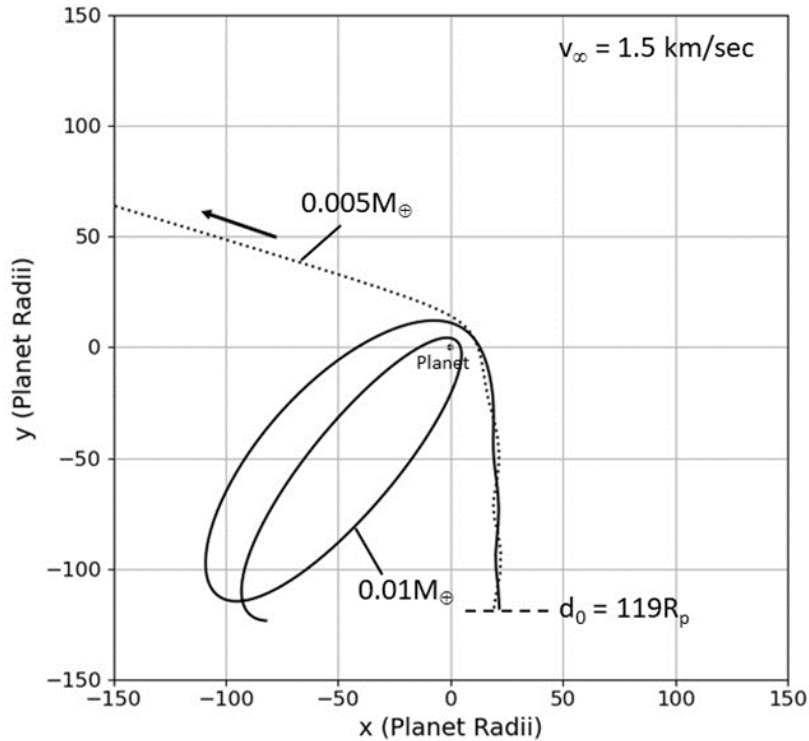

**Fig. 4:** Encounter involving the same binary as in Fig. 3 but started at a slightly greater distance d = 100 $R_\oplus$ + 19 $R_\oplus$ = 119 $R_\oplus$ below the planet. The captured mass $m_1 = 0.01$ $M_\oplus$ and the escaping mass $m_2 = 0.005$ $M_\oplus$. The resulting capture orbit has $q = 2.57$ $R_p$, $a = 77.8$ $R_p$, and $e = 0.966$ after the second pass by the planet.

Now the larger mass is shown to be captured in Fig. 4 but on an orbit with greater $a$ and $e$. This is because the larger mass has a smaller circumbinary velocity that contributes a smaller deceleration in the planet reference frame. Fig. 5 compares the velocity in the planet frame of both masses leading to capture of the smaller mass (Fig. 5A) and capture of the larger mass (Fig. 5B). In both cases, capture occurs when velocity dips below the escape velocity near the moment of disruption. A comparison of the post-capture velocities reveals that the lighter mass is moving slower in Fig. 5A than the heavier mass in Fig. 5B owing to the initial differences in circumbinary velocities before tidal disruption. The greater velocity of the heavier mass post-capture is responsible for the distended elliptical orbit in Fig. 4.

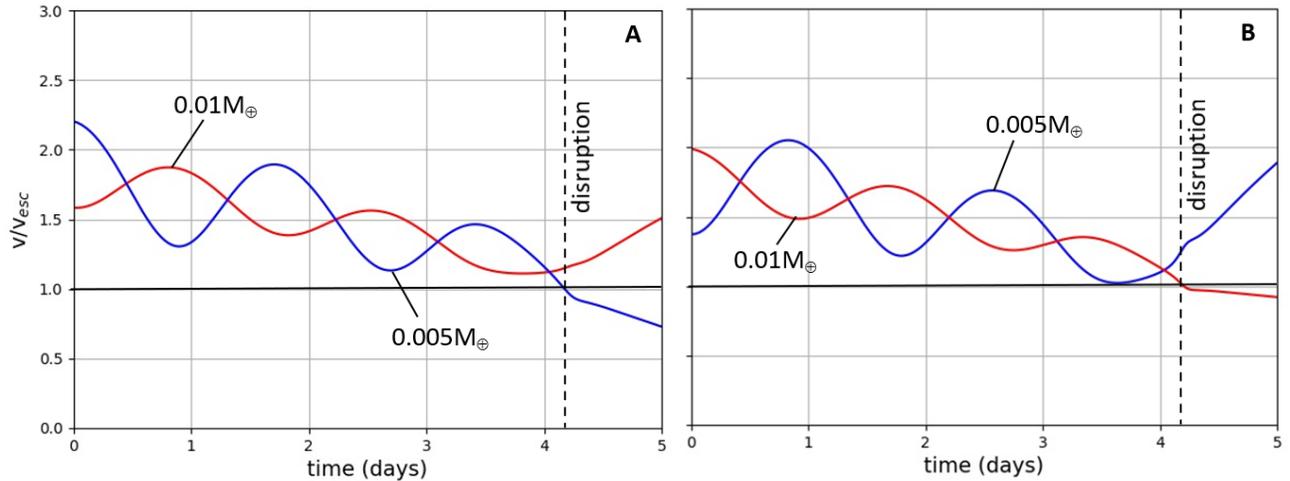

**Fig. 5:** Velocities of the binary masses (red for 0.01 $M_\oplus$ and blue for 0.005 $M_\oplus$) for the encounters shown in Fig. 3 (panel **A**) and Fig. 4 (panel **B**). Velocities are computed in the planet reference frame and scaled by the planet escape velocity. Dashed lines mark the approximate time of tidal disruption.

Long-term stability of the capture orbit, especially for the larger mass in Fig. 4, is in question. Careful analysis reveals that apoapse distance $a(1+e) \approx 154\ R_p$ in Fig. 4 is greater than the stability limit ~0.5 $r_H \approx 118\ R_p$ for a prograde orbit around Earth at 1.0 AU from the Sun, which implies that the orbit will be short-lived. Strong solar perturbations are evident in this case, causing the orbit to flex and precess significantly in only two orbital cycles. We examine the question of long-term stability of eccentric satellite orbits in the next section of the paper.

Here we return our attention to retrograde capture, which possibly occurred with Triton around Neptune. We stated earlier that retrograde orbits are more stable than prograde and that retrograde-moving satellites are slightly easier to capture than prograde. We tested this numerically by starting a binary *inside* the planet's orbit (see Fig. 6) rather than *outside* as in Figs. 3-4. We chose masses and a starting velocity for the binary that fit with retrograde capture according to Fig. 2B. However, we also started the binary rotating counterclockwise, the same as in Figs. 3-4 but now with *spin* and *orbital* angular momenta (around the planet) of the binary in opposite directions. This key difference opposes both tidal disruption and capture. First, coriolis forces within the moving binary are reversed which opposes separation. Second, the *escaping* mass is now the *inner* mass (see Fig. 1) which interferes with and prevents capture of the *outer* mass, even if the outer mass is moving under the escape velocity around the planet. This scenario is illustrated graphically in Fig. 6A, and with the successful retrograde capture in Fig. 6B.

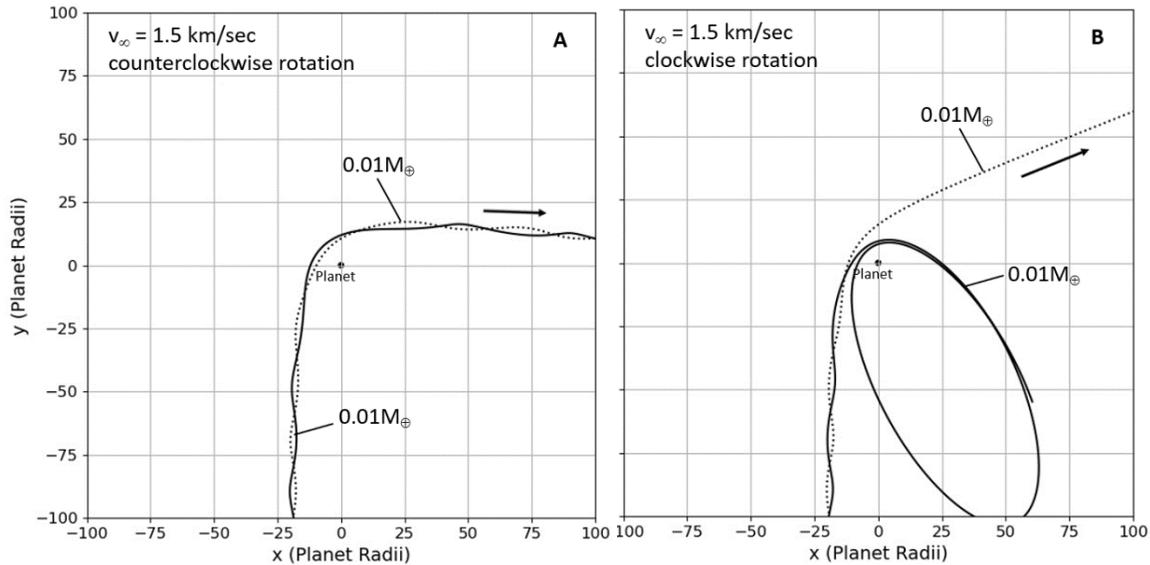

**Fig. 6:** A retrograde encounter involving an equal-mass binary (m = 0.01 M⊕) approaching an 'earth' from the left. In the heliocentric frame, the planet is moving vertically and the Sun is offscreen to the left. Disruption and capture does not occur in **A** with the binary rotating counterclockwise but does occur in **B** with the binary rotating in the opposite direction.

Therefore, retrograde capture of satellites through binary-exchange is most favorable when the spin- and orbital-angular momenta of the encountering binary are in the same direction.

We conclude this section by investigating how large of a satellite can be captured around an Earth-mass planet. The curves in Fig. 2 indicate that prograde capture of Mars-size satellite should be possible if the escaping mass $m_2 > 0.02$ M⊕. We tested this by inserting a 'mars' as one of the binary members, and again using a small approach velocity ($v_\infty = 1.5$ km sec$^{-1}$) to promote capture. We then integrated the encounters repeatedly by adjusting the smaller binary mass and initial x-y position of the binary to capture the largest mass. This simplistic Monte-Carlo approach involved ~30 runs, each consuming ~2 minutes of CPU time on a PC, and yielded a few notable captures. Two are shown in Fig. 7 below. Both encounters result in capture of a Mercury-sized satellite around an 'earth' made possible by the escape of second larger object the size of Mars. The captured orbits are shown to be tight ($a < 25$ $R_p$) and stable, albeit with sizeable eccentricities.

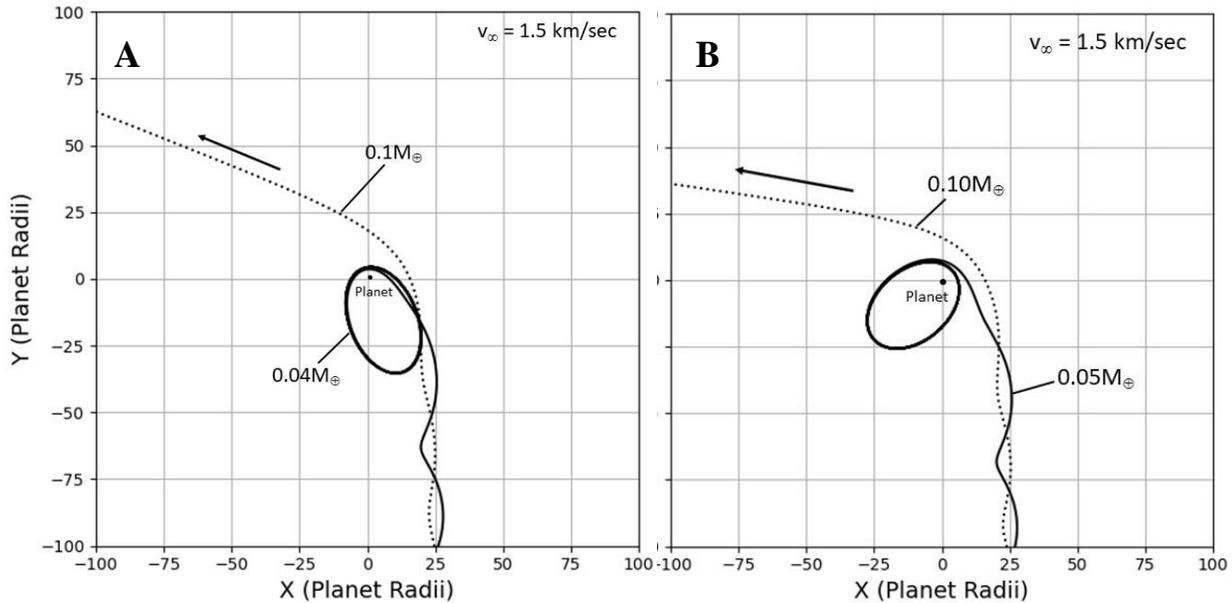

**Fig. 7**: Encounters involving an Earth-mass planet and binaries containing a Mars-sized mass (m = 0.1 $M_⊕$), which escapes, and smaller captured objects having m = 0.04 $M_⊕$ (panel **A**) and m = 0.05 $M_⊕$ (panel **B**). Distance from the Sun (1 AU) and encounter velocities at infinity ($v_∞$ = 1.5 km sec$^{-1}$) are the same for both runs. The capture orbit in panel **A** has $q$ = 3.68 $R_p$, $a$ = 21.1 $R_p$, and $e$ = 0.825, and the capture orbit in panel **B** has $q$ = 5.21 $R_p$, $a$ = 20.3 $R_p$, and $e$ = 0.743.

Upon demonstrating that capture of a 'mercury' around 'earth' is possible, we methodically adjusted the starting location of the binary so that the larger Mars-sized mass was appropriately positioned for capture at the moment of disruption. After several attempts, we were able to find a case where the planet-centric velocity of 'mars' dipped below the escape velocity of the planet ($v/v_{esc}$ ~0.95), resulting in a temporary capture shown in Fig. 8. However, the sizeable apoapse distance of the capture orbit, $a(1+e)$ = 114 $R_p$, is near the edge of the orbital stability limit around 'earth' at 1.0 AU from the Sun, the same as for the captured object in Fig. 4. Thus, the orbit will likely be similarly short-lived, although we do not examine its long-term evolution here.

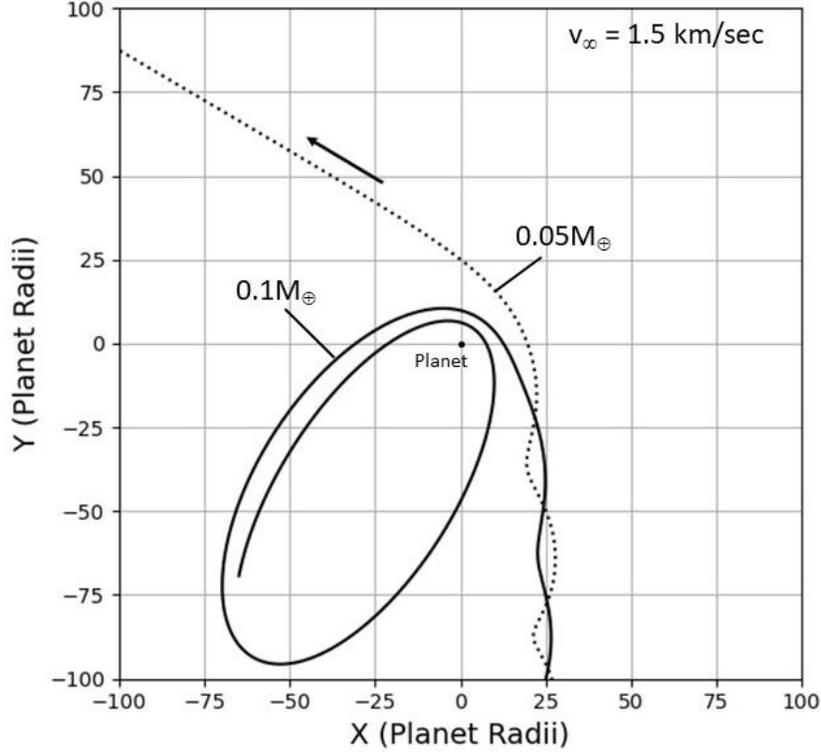

**Fig. 8**: Fragile capture of a Mars-sized mass (m = 0.1 $M_\oplus$) around an 'earth' with run parameters the same as in Fig. 7B except with starting distance of the binary increased by 10 $R_p$. The capture orbit has $q = 5.85$ $R_p$, $a = 60.0$ $R_p$, and $e = 0.902$. The orbital path is shown for $t = 0.1$ years, and substantial apsidal precession of the orbit is evident from strong solar perturbations.

The example in Fig. 8 raises a key question regarding the application of the analytic capture limits determined by Eq. (1) and the curves in Figs. 2-3. According to Fig. 2, an escaping mass $m_2 \sim 0.05$ $M_\oplus$ should permit capture of an object $> 0.1$ $M_\oplus$, even though we were unable to numerically capture anything larger than this. The reason for this discrepancy stems from the original derivation of Eq. (1) in W13, which assumes the binary masses to be small compared to the planet. But post-disruption orbital velocity $v$ of $m_1$ around $M_p$, with separation of masses held constant, scales as

$$v \propto M_p/(M_p + m_1)^{1/2}$$

which decreases as $m_1/M_p$ increases. This means that the captured mass must lose more energy in the *exchange* to achieve the same orbit around a planet of lesser mass, thereby making the maneuver more difficult and less likely. Thus, the accuracy of the analytic result of Eq. (1) diminishes as the ratio of the binary masses to the planet mass increases, leaving numerical integration as the most reliable tool for analyzing binary-exchange capture in these extreme cases.

## 4. Post-Capture Tidal Evolution

We have shown in the last section that large terrestrial masses (> 0.05 $M_\oplus$) might be captured around Earth-sized planets and inserted into orbits of significant eccentricity (*e* > 0.6). Whether such objects remain as satellites for Gyr timescales depends on an assortment of physical and dynamic parameters of the host planets and their new satellites, as well as distance of the host star. Here we examine the short-term tidal evolution of a candidate eccentric planet-satellite system – an 'earth' with a lunar-sized satellite - to test whether binary-exchange might apply to the Earth-Moon system, and by plausible extension, to other massive satellites that may have once existed in the inner Solar System (Burns 1973).

Figure 2 shows that Earth could possibly capture the Moon (~0.012 $M_\oplus$) if the Moon originates within a binary having a companion mass >0.005 $M_\oplus$. While no binary of this size exists within 40-50 AU of the Sun today (Brown et al. 2006), the early inner Solar System might have harbored some massive binaries, possibly formed through collision, as might have happened with Pluto and Charon (Canup 2005; Cheng et al. 2014). Such an early collision might have depleted the Moon in both core metals and volatiles, two things that are needed to explain the low lunar density and composition of lunar rocks (Barr 2016; Cano, E.J. *et al* 2020).

The fate of any satellite is determined by its proximity to its host planet compared to the *co-rotation* (or 'corot') radii of both objects, which is the distance at which the orbit- and spin-rates are equal. The early spin period $P_{rot}$ of an isolated Earth is unknown but is generally assumed to be ~5-10 hours owing to the violence of the largest impacts in the final stages of accretion (Dones and Tremaine 1993). We arbitrarily set the initial planet spin period $P_{rot}$ = 7 hours for all tidal calculations that follow, which corresponds to a corot radius of 2.93 $R_p$ around Earth.

The spin rate of a hypothetical captured object during binary-exchange would be much slower, as it will likely be tidally synchronized with the orbital period of the binary, which is related to the assumed orbital separation $a_B = 10(R_1 + R_2)$ through Kepler's third law. For a terrestrial density $\rho$ = 4000 kg m$^{-3}$ and mass in the range *m* = 0.01-0.1 $M_\oplus$, typical binary orbital periods are ~days. We, therefore, set the initial spin period of the captured object to $P_{rot}$ = 2 days under the assumption of synchronized spin before tidal separation during an encounter. This spin period corresponds to a corot radius of 14.0 $R_m$ for a lunar-sized mass ($R_m$ = 1738 km), which places the planet involved in binary-exchange inside the corot radius of the satellite near periapse in a small orbit of significant eccentricity.

This situation is depicted in Fig. 9A where the tide raised by the planet (the Earth) on the satellite (the Moon) lags slightly behind the line connecting the object centers. This tiny time lag causes the Moon's spin rate to increase and the Earth-Moon separation to decrease. The tide raised by the Moon on the Earth opposes this change, but the strength of the "orbit-shrinking" tide ($\propto M^2 R^5$, where $M$ is the mass of the perturbing object and $R$ is the radius of the distorted object) is ~9.8 times that of the "orbit-expanding" tide, and the Moon moves toward the Earth so long as the Earth remains inside the corot radius. In Fig. 9B, the spin rate of the Moon has increased (as shown in the calculation in *Section 4.1*) after $t = 100$ years so that neither object is within the other's corot radius, and the tidal evolution reverses direction causing the orbit to expand.

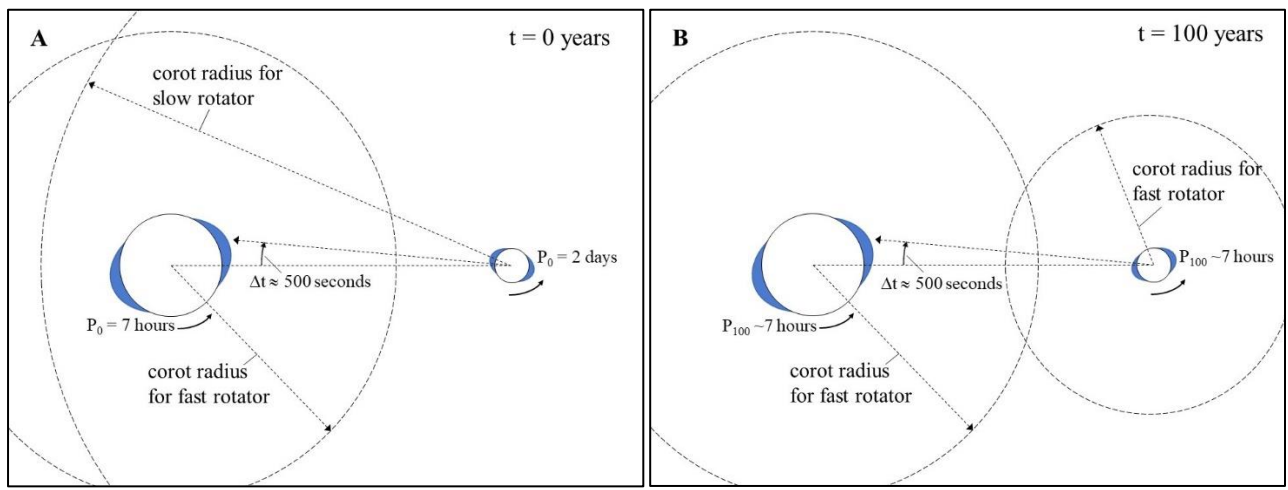

**Fig. 9**: Relationship of tidal-bulge orientation to object spin rate and the corot radius. Objects in both panels are the Earth on the left and the Moon on the right, although distance and size are not to scale. The frames depict snapshots of the system near periapsis in an eccentric orbit when tidal accelerations are maximum. Tidal bulges are shown as shaded blue extensions to the circular disks of the planet and satellite. Maximum tide is shown to lag/precede the line of symmetry by a time $\Delta t$ for both objects in the constant time-lag model described below. Panel **A** shows conditions at the start of the tidal calculation, and panel **B** is a century later, when the Moon's spin rate has increased considerably, and Earth is no longer inside the Moon's corot radius when near periapsis in an eccentric orbit.

*4.1 Constant Time-Lag Tidal Model*

Our strategy is to numerically integrate the tidal evolution of a captured terrestrial satellite around the Earth with a large primordially eccentricity. We employ a version of the constant time-lag tidal model first developed by Mignard (1980) and Hut (1981) but more recently utilized by several authors (Leconte et al. 2010, Nogueira 2011, Cheng et al. 2014). We refer the studious reader to Eqns. (3-7) of Cheng et al., where expressions for determining rates of orbital change (*da/dt*, *de/dt*) and spin evolution are clearly written. We apply those expressions to a planet-satellite system containing the Earth and Moon with tidal parameters listed in Table 1 below.

**Table 1**: Model parameters used for tidal calculations

| Initial spin period for planet | 7 h |
|---|---|
| Initial spin period for captured satellite | 2 d |
| Tidal Love number *k* for both | 0.1 |
| Constant tidal time lag *Δt* for both | 500 s |
| Obliquity of spin axis for both | 0 deg |

Tidal parameters *k* and *Δt* above are interpreted as standard values for an assortment of terrestrial objects. The actual Love number for the Moon today is thought to be smaller: $k = 0.02664$ (Zhang 1992). And the often-quoted value for Earth is larger ($k \sim 0.3$ according to Stacey and Davis 2008), owing to the small rigidity of the oceans. The value of $\Delta t \propto Q^{-1}$ is a measure of the tidal energy dissipated in a rotational cycle, where the dissipation parameter $Q \sim 13$ for the Earth (Burns 1973) and $Q \sim 27$ for the Moon (Yoder 1995), although these values do change with time.

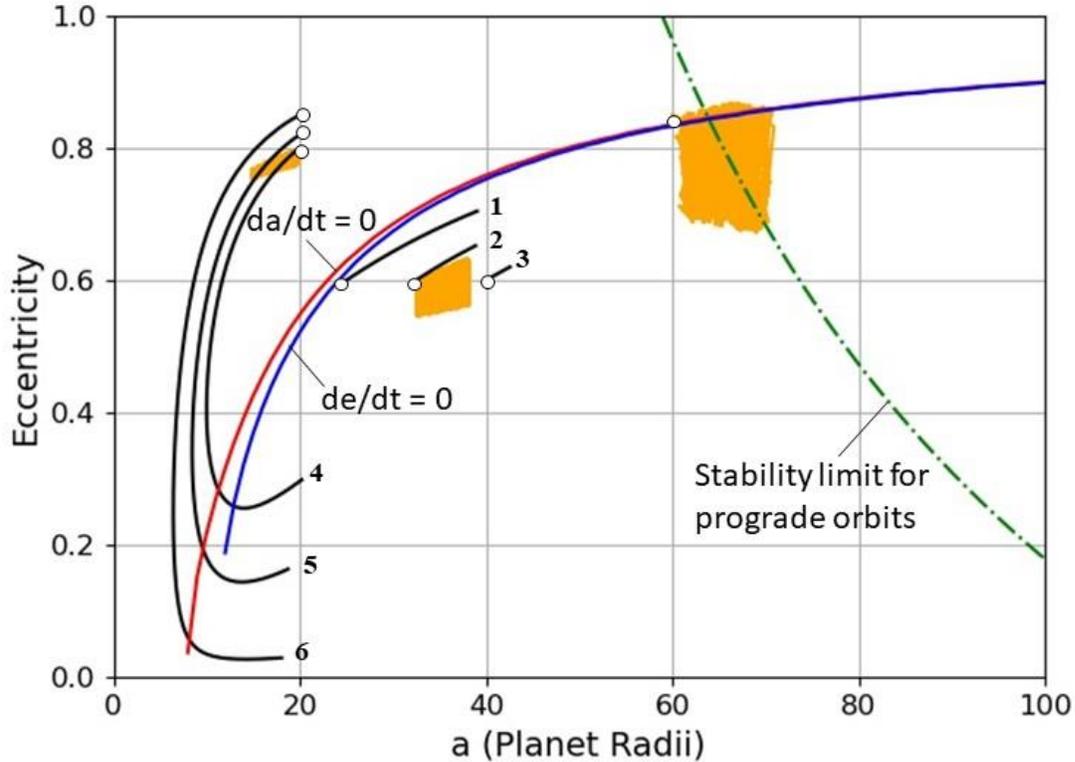

**Fig. 10**: Tidal changes (solid black lines) in orbit eccentricity and semi-major axis over 1Myr for the Moon around the Earth with different starting values of *a* and *e* indicated with open white circles. Red and blue arcs indicate where *da/dt* and *de/dt* equal zero. The stability limit for prograde orbits (green dot-dashed line) is calculated using Eq. (3) with $f_H = 0.5$ and with *a* = 1AU. Orange patches show high-frequency eccentricity "Kozai" (Kozai 1962) oscillations that appear as solid color, and calculated using *REBOUNDx* which includes acceleration by the Sun. Orbital parameters at the start of the *REBOUNDx* integrations are also indicated with open white circles.

Close inspection of the orbit-averaged Eqns. (3-5) of Cheng. et al. (2014) show that the direction of tidal evolution is solely determined by the sign of the terms in square brackets [], which varies with the ratio of spin frequency to orbit frequency (which depends on *a*) of both objects, as well as the eccentricity *e*. We calculate the values of *a* and *e* where *da/dt* and *de/dt* = 0 (that is, the brackets [] in Eqns. (4) and (5) of Cheng et al. = 0) and plot these stationary lines as parabolic arcs in an *e* vs *a* diagram in Fig. 10 above. The location and shape of the *da/dt* and *de/dt* arcs are nearly identical, which conveniently delineates a numerical boundary between orbit expansion (below the lines) and contraction (above).

We numerically solve the orbit-averaged Eqns. (3-5) of Cheng. et al. using a simple Euler technique and a small step size $dt = 0.1$ years, with obliquity assumed to be zero for both objects and with the Sun at 1.0 AU included only in the determination of object spin. The 1 Myr tidal-evolution curves in Fig. 10 took ~15 min of CPU time on a standard PC to compute and are shown for six runs with starting parameters listed in Table 2 below.

**Table 2**: Initial orbital parameters used for tidal calculations in Fig. 10

| Run | $a$ ($R_p$) | $e$ |
|---|---|---|
| 1 | 25.0 | 0.60 |
| 2 | 32.5 | 0.60 |
| 3 | 40.0 | 0.60 |
| 4 | 20.0 | 0.80 |
| 5 | 20.0 | 0.82 |
| 6 | 20.0 | 0.85 |

Although acceleration of the system by the Sun was not included in the original six runs, we followed up by testing our Euler algorithm against the open-source code *REBOUNDx*, and specifically the "modify_orbits_forces" routine (Kostov et al. 2016) which calculates orbits in the usual way using the extremely accurate *IAS15* integrator (Rein and Spiegel 2015) while adjusting orbit semi-major axis and eccentricity according to:

$$a = a_0\, e^{t/\tau_a},$$
$$e = e_0\, e^{t/\tau_e}, \quad (5)$$

with exponential growth parameters $\tau_a$ and $\tau_e$ given by:

$$\tau_a = \left(\frac{1}{a}\frac{da}{dt}\right)^{-1},$$
$$\tau_e = \left(\frac{1}{e}\frac{de}{dt}\right)^{-1}, \quad (6)$$

which are easily computed from the tidal Eqns. (4) and (5) of Cheng et al. (2014). We used these expressions to update $a$ and $e$ and the planet-satellite spin rates every 0.1% of an integration, which lasted 1 Myr for *REBOUNDx* runs starting from $a = 32$ $R_p$ (run "32") and 60 $R_p$ (run "60"), and 200 years for the run starting from $a = 20$ $R_p$ (run "20"). *REBOUNDx* runs each took ~90-min of CPU time to compute, including the shortest run #20 since the satellite distance and *IAS15* step-size were smaller.

*REBOUNDx* run #20 was not only the most expensive (years of integration/CPU minute), but it also showed the greatest difference between the two integration strategies, starting from the same open circle at $a = 20$ $R_p$ and $e = 0.8$ in Fig. 10. This discrepancy likely stems from differences in integration step size and tidal update frequency near periapsis at high eccentricity. Nevertheless, the trend toward tidal circularization starting from high eccentricity is evident and similar to Euler runs #4-6.

Tidal expansion is evident in Euler runs #1-3 and the outer two *REBOUNDx* runs #32 and #60. In the *e*-vs-*a* diagram, the rate of tidal change diminishes rapidly toward the right with increasing *a*, falling as $a^{-7}$ for *da/dt* and $a^{-8}$ for *de/dt*. This explains why lengths of tidal tracks #1-3 decrease with increasing distance from the planet. Also, run #32 closely resembles Euler run #2 and may, if followed for an additional 5-10 Myr, trend toward instability. *REBOUNDx* run #60 exhibits the greatest variability as it straddles the outer stability limit and is at greatest risk of satellite escape in < 10 Myr.

Our purpose here is to demonstrate the difference between tidal expansion and contraction and which regions of Fig. 10 yield rapid significant change capable of preserving a captured satellite for Gyr timescales. Orbit expansion is permissible so long as the tidal evolution is slow, as it has been for the Moon around the Earth for most of its history. This is the case for satellites on orbits of small eccentricity near the bottom of Fig. 10.

Tidal rates increase moving upward in the plot, reaching levels that are several orders of magnitude higher when $e > 0.8$, compared to $e < 0.1$. These extraordinary rates stem from extreme pulses of tidal acceleration near periapsis at high eccentricity. With $a = 20$ $R_p$ and $e = 0.85$ in run #6, for example, periapsis distance $q$ is only 3 $R_p$, where both the planet and satellite are within the partner corot radius (albeit for < 5% of the orbit), causing the orbit to rapidly contract and circularize in under 100 years; see Figures 11 and 12. For comparison, the Roche distance from the Earth for tidal destruction of the Moon on a circular orbit is ~2.9 $R_p$. Thus, a captured mass should be able to weather the tidal extremes experienced over a tiny fraction of the orbital period.

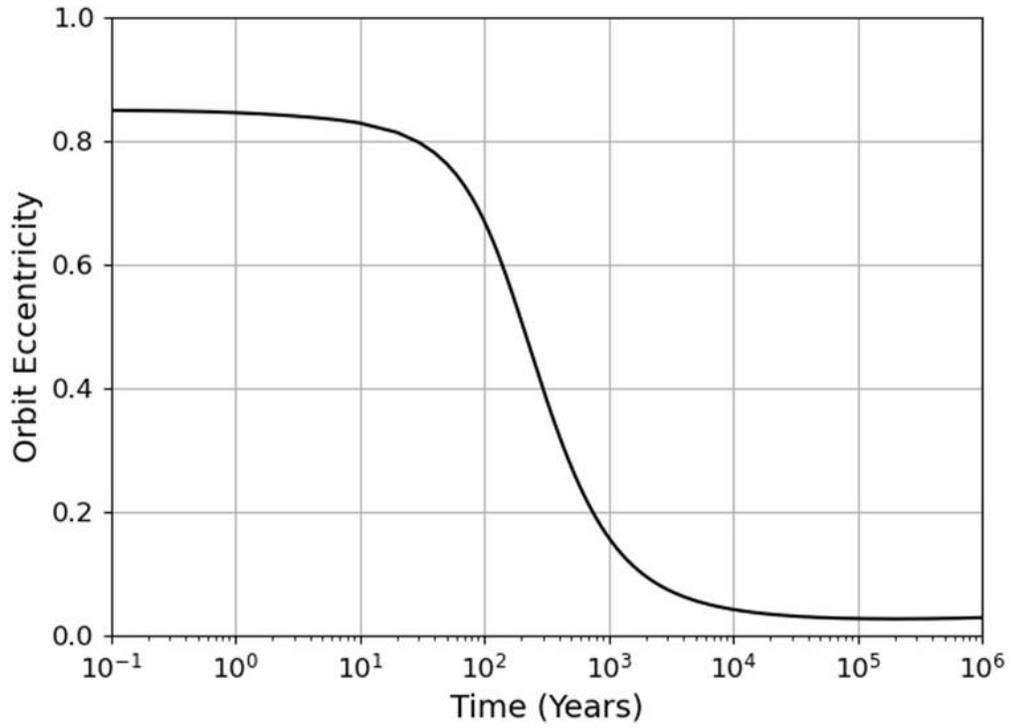

**Fig. 11**: Rapid tidal circularization in run #6 shown in Fig. 10.

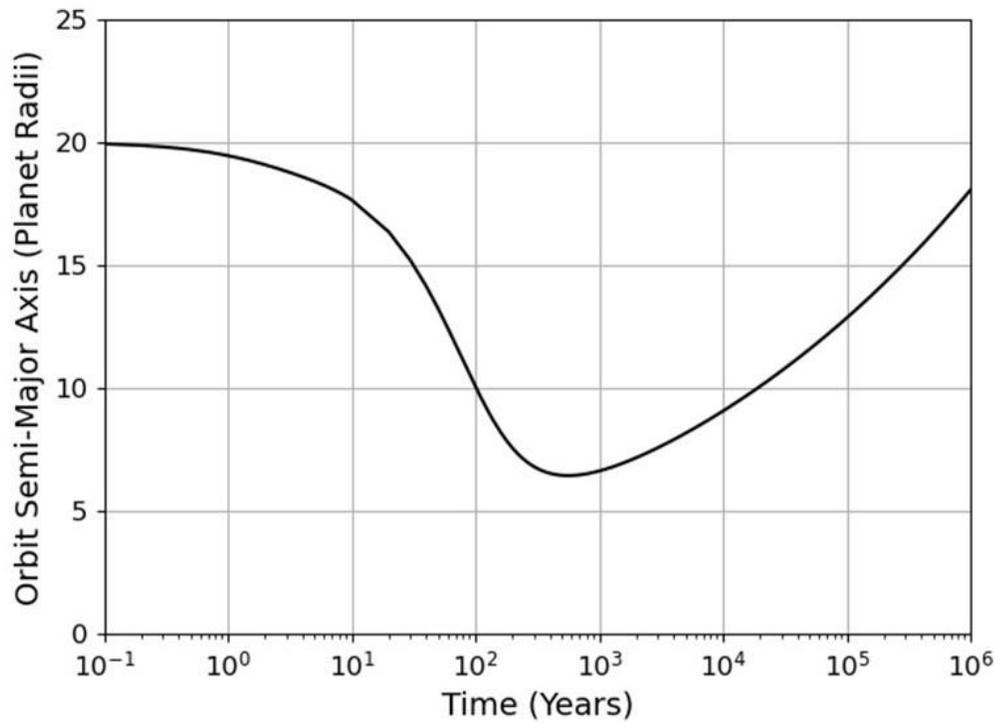

**Fig. 12**: Rapid tidal contraction followed by slow expansion in run #6 shown in Fig. 10.

Orbit contraction and circularization is accompanied by acceleration of satellite spin as explained above and depicted in Fig. 9. This spin-up converts the satellite from a slow-rotator post capture to a fast rotator, which shrinks its corot radius and reverses the orientation and effect of the planet-induced tide (Fig. 9b). The orbit is then made to expand (Fig. 12), albeit far more slowly with low eccentricity, at the expense of planet and satellite spin (Fig. 13).

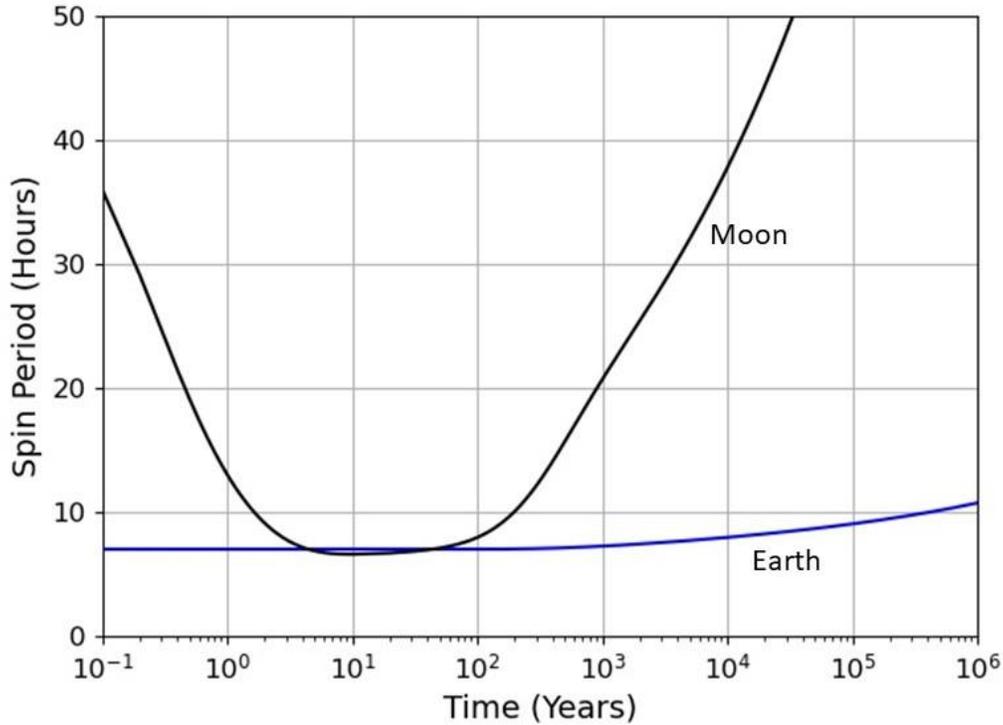

**Fig. 13**: Tidal changes to spin rate for the satellite (the Moon) and the planet (Earth) in run #6 shown in Fig. 10.

---

Captured satellites will sport eccentricities in the range $0.6 < e < 0.9$. According to Fig. 10, most orbits in this category will expand and are subject to significant solar perturbation. However, Fig. 10 also shows that objects with high $e$ and small $a$ (< ~20 $R_p$) will have their orbits rapidly circularized by tides in under 1000 years, with the amount of circularization determined by initial $a$ and $e$. Tides also cause orbits to contract and satellites to spin faster (assuming slow synchronous rotation within the binary before disruption), which initiates a negative feedback that counters the inward evolution. In this way, a captured satellite originally on an elliptical trajectory is converted into a semi-permanent 'moon' (possibly *the* Moon) on an approximately circular orbit (Fig. 10; run #6) and trapped in a long tidal drift away from the planet lasting billions of years.

Considering the effect of tidal contraction and circularization on Earth's rotation, the spin rate is only weakly affected with Earth's day lengthening from $P = 7$ hours to $P \sim 10$ hours during the first 1 Myr of subsequent tidal drift (Fig. 13). The distance of the Moon after this short expansion is $\sim 18$ $R_p$ according to Fig. 12, which is compatible with early tidal parameters of the system (Goldreich 1966) needed to explain the present lunar distance and spin rate of the Earth.

## 5. Discussion

We have shown that binary-exchange capture works just as well for terrestrial planets as it does for gas-giant planets (W13). Capture of objects in the mass range 0.01–0.1 $M_\oplus$ by an 'earth' around the Sun or another star is in principle possible, even though it is not yet tested whether binary-exchange does occur with measurable frequency within a realistic, three-dimensional swarm of planetesimals with randomized binary orientations.

However improbable these events may be, such a capture may have already occurred in the outer Solar System to form Triton around Neptune (Agnor and Hamilton 2006). It has also long been argued that collisional formation of binaries has occurred in both the inner and outer Solar System, possibly forming the Moon around the Earth (see Barr 2016 and the references therein). So, the existence of BTOs in the early Solar System is not a hard sell, even though their genesis through N-body simulation has yet to be studied with care.

Whether the above scenario applies to the Earth and Moon is an open question. A weakness of the idea is that it requires two improbable events - a collision followed by a capture – whereas formation through collision (i.e., a "giant impact") only requires one. Setting aside this debate for now, we consider that the Moon may have first belonged to a terrestrial binary having a companion mass $m > 0.005$ $M_\oplus$ originally formed through collision. This early collision may have depleted lunar material in volatiles and heavy elements by an amount similar to if it had formed around an impacted Earth. However, such a collision may require a companion mass significantly larger than the Moon to yield the small size for the lunar core; Pluto and Charon are presumed to have formed through collision (Canup 2005) and yet sport similar densities and core sizes. This deserves additional study through SPH integration.

Once a binary is formed, the encounter velocity leading to capture must be slow, with $v_\infty < 3$ km sec$^{-1}$. This favors encounters involving masses that are in close proximity within a given terrestrial-mass population. Thus, the binary cannot originate more than 0.1-0.2 AU from the Earth if the orbit intersection velocity is to be less than this. This means that a captured moon would have originated in the same orbital annulus as the Earth ~1 AU from the Sun, which might explain the similar, yet distinct, isotopic compositions of lunar and terrestrial rocks (Cano, E.J. *et al* 2020, Dauphas, N. et al. 2014, Lock, S.J. et al. 2018).

Lastly, since the inclination of the capture orbit is randomly determined by the angular momentum of the spinning binary with respect to the planet-centric encounter direction, there is a wide range of possible inclinations for a Moon formed through binary-exchange, just as with Triton and Neptune. Thus, there are no intractable constraints imposed by the current ~5° inclination (wrt the ecliptic plane) of the lunar orbit, which is otherwise difficult – though, not impossible (Touma and Wisdom 1994a; Cuk, M. et al. 2021) - to explain. This represents a significant advantage over existing models of lunar formation.

## 6. Summary

We have examined the concept of collision-less binary-exchange for capturing massive satellites - comparable to and larger than the Moon - around terrestrial-mass objects either inside or outside the Solar System. We use an analytical approximation to outline the physics of the interaction and to set limits on the masses that can possibly be captured. Capture through binary-exchange favors low-mass binaries (relative to the size of planet) traveling at small relative velocity (< 3 km/sec) and rotating in the same direction as their angular momentum around the planet. A tidal model is then applied to an 'earth' with a lunar-mass satellite on a post-capture orbit with a significant eccentricity. Orbit contraction and circularization are possible if the satellite experiences rapid tidal damping, which leads to long-term orbital stability, as is the case for the Earth and Moon. We conclude that binary-exchange capture is a viable process in forming large satellites in the inner-regions of solar-type planetary systems.

**Acknowledgements:**

We thank Andrew Rosenswie for a helpful review.

**References:**

Agnor, C.B. Hamilton., D.P. 2006. Neptune's capture of its moon Triton in a binary planet gravitational encounter. *Nature* **441**, 7090.

Barr, A.C. 2016. On the origin of the Earth's moon. *Journal of Geophysical Research: Planets* **121,** 1573–1601.

Brown, M.E. *et al* 2006. Satellites of the largest Kuiper Belt objects. *Astrophys. J.* **639** L43.

Burns, J.A. 1973. Where are the satellites of the inner planets? *Nature Physical Science* **242**, 23–25.

Cano, E.J. *et al* 2020. Distinct oxygen isotope compositions of the Earth and Moon. *Nature Geoscience* **13**, 270–274.

Canup, R.M. 2005. A giant impact origin of Pluto-Charon. *Science* **307**, 5709.

Canup, R.M., Esposito, L.W. 1996. Accretion of the Moon from an Impact-Generated Disk. *Icarus* **119.2**, pp. 427-446.

Canup, R.M., Ward, W.R. 2002. Formation of the Galilean Satellites: Conditions for Accretion. *Astron. J.* **124**, pp. 3404-3423.

Chambers, J.E. 2013. Late-stage planetary accretion including hit-and-run collisions and fragmentation. *Icarus*, **224,** 43–56.

Cheng, W.H., Lee, M.H., Peale, S.J. 2014. Complete tidal evolution of Pluto-Charon. *Icarus*, 233, 242–258.

Ćuk, M., Lock, S.J., Stewart, S.T., Hamilton, D.P. 2021. Tidal evolution of the Earth-Moon system with high obliquity. *Planetary Science Journal*, **2:147**, 1–11.

Dauphas, N. et al. 2014.Geochemical arguments for an Earth-like Moon-forming impactor. *Phil. Trans. R. Soc. A*. **372**

Dones, L., Tremaine, S. 1993. Why does Earth spin forward? *Science* **259**, 350–354.

Estrada, P.R., Mosqueira, I. 2006. A gas-poor planetesimal capture model for the formation of giant planet satellite systems. *Icarus* **181**, 486-509.

Goldreich, P. 1966. History of the lunar orbit. *Rev. Geophys.*, **4(4)**, 411–439.

Goldreich, P, Murray N., Longaretti, P.Y., Banfieldet, D. 1989. Neptune's story. *Science* **245(4917)**, 500-504.

Grishin, E., Malamud, U., Perets, H.B. et al. 2020. The wide-binary origin of (2014) MU69-like Kuiper belt contact binaries. *Nature* **580**, 463–466.

Hut, P. 1981. Tidal evolution in close binary systems. *Astron & Astrophys*. 99, 126–140.

Kozai, Y. 1962. Secular perturbations of asteroids with high inclination and eccentricity. *Astron J.* **67**, 591.

Kostov, V.B., Moore, K., Tamayo, D., Jayawardhana, R., Rinehart, S.A. 2016. Tatooine's Future: The Eccentric Response of Kepler's Circumbinary Planets to Common-envelope Evolution of Their Host Stars. *Astrophys. J.* **832**, 183.

Leconte, J., Baraffe, I., Chabrier, G., Levrard, B. 2010. Is tidal heating sufficient to explain bloated exoplanets? Consistent calculations accounting for finite initial eccentricity. *Astron & Astrophys* **516**, A64.

Lock, S.J., Stewart, S.T., Petaev, M.I., Leinhardt, Z., Mace, M.T., Jacobsen, S.B., & Ćuk, M. 2018. The origin of the Moon within a terrestrial synestia. *Journal of Geophysical Research: Planets* **123**, 910– 951.

Margot J.-L., Pravec P., Taylor P., Carry B., and Jacobson S. 2015. Asteroid systems: Binaries, triples, and pairs. In Asteroids IV (P. Michel et al., eds.), pp. 355–374. Univ. of Arizona, Tucson

Mignard, F. 1980. The evolution of the lunar orbit revisited, II. *The Moon and the Planets* **23**, 185–201.

Morbidelli, A., Levison, H.F., Bottke, W.F., Dones, L., Nesvorn´y, D. 2009. Considerations on the magnitude distributions of the Kuiper belt and of the Jupiter Trojans. *Icarus* **202**, 310-315.

Nogueira, E., Brasser, R., Gomes, R. 2011. Reassessing the origin of Triton. *Icarus* **214.1**, 113-130.

Noll, K.S. 2005. Solar system binaries. *International Astronomical Union.Proceedings of the International Astronomical Union.* **1**, 301-318.

Noll, K.S., Grundy, W.M., Chiang, E.I., Margot, J.-L., Kern, S.D. 2008. Binaries in the Kuiper Belt. The Solar System Beyond Neptune. 345-363. M.A. Barucchi, H. Boehnhardt, D.P. Cruikshank, A. Morbidelli (eds). University of Arizona Press, Tucson, AZ, USA.

Noll, K.S., Grundy, W.M., Nesvorný, D. and Thirouin, A., 2020. Trans-neptunian binaries (2018). In *The trans-Neptunian solar system* (pp. 205-224). Elsevier.

Porter, S.B., Grundy, W.M. 2011.Post-capture Evolution of Potentially Habitable Exomoons. *Astrophysical Journal* **736.1** L14.

Raymond, S.N., O'Brien, D.P. 2009. Building the terrestrial planets: Constrained accretion in the inner Solar System. *Icarus* **203**, 664–662.

Rein, H., Spiegel, D.S. 2015. IAS15: a fast, adaptive, high-order integrator for gravitational dynamics, accurate to machine precision over a billion orbits. *MNRAS* **446**, 1424–1437.

Ronnet, T., Johansen, A. 2020. Formation of moon systems around giant planets - Capture and ablation of planetesimals as foundation for a pebble accretion scenario *A&A* **633** A93

Rufu, R. and Canup, R.M. 2017. Triton's Evolution with a Primordial Neptunian Satellite System. *Astron. J.* **154**, 208-216.

Stacey, F., Davis, P. 2008. Physics of the Earth (4th ed.). Cambridge: Cambridge University

Touma, J., & Wisdom, J. 1994a, *Astron. J.*, **107**, 1189

Walsh, K.J., & Jacobson, S.A. 2015. Formation and Evolution of Binary Asteroids. In Asteroids IV, Patrick Michel, Francesca E. DeMeo, and William F. Bottke (eds.), University of Arizona Press, Tucson

Williams, D.M. 2013. Capture of Terrestrial-Sized Moons by Gas-Giant Planets. *AstroBio* **13.4**, 315-323.

Yoder, C.F. 1995. Astrometric and Geodetic Properties of Earth and the Solar System. In Global Earth Physics, T.J. Ahrens (Ed.).

Zhang, C.Z. 1992. Love numbers of the moon and of the terrestrial planets. *Earth Moon Planet* **56**, 193–207.